\documentclass{ws-ijmpcs}

\usepackage{graphicx}
\usepackage{epstopdf}
\usepackage{bm}
\usepackage{epsfig}
\usepackage{graphics}
\usepackage{xspace}
\usepackage{subfigure}

\def\trimmarks{}

\def\OMIT#1{}

\newcommand{\nn}{\nonumber}

\newcommand{\bea}{\begin{eqnarray}}
\newcommand{\eea}{\end{eqnarray}}

\newcommand{\gsim}{\mathrel{\rlap{\lower4pt\hbox{\hskip1pt$\sim$}}\raise1pt\hbox{$>$}}}

\newcommand{\be}{\begin{equation}}
\newcommand{\ee}{\end{equation}}

\begin{document}

\markboth{Z.Kang,X.Liu,S.Mantry,J.Qiu}
{1-jettiness Event Shape for DIS Processes}

%
\catchline{}{}{}{}{}
%

\title{THE 1-JETTINESS EVENT-SHAPE FOR DIS \\WITH NNLL RESUMMATION}
\author{ZHONG-BO KANG}

\address{Los Alamos National Laboratory,
                   Theoretical Division,
                   Los Alamos, NM 87545, USA}

\author{XIAOHUI LIU}

\address{High Energy Division, 
                  Argonne National Laboratory, 
                  Argonne, IL 60439,USA\\                 
                  Department of Physics and Astronomy, 
                   Northwestern University,
                   Evanston, IL 60208,USA}

\author{SONNY MANTRY\footnote{Talk presented at the QCD Evolution Workshop, JLab, May 6th-10th, 2013.}}

\address{High Energy Division, 
                  Argonne National Laboratory, 
                  Argonne, IL 60439,USA\\                 
                  Department of Physics and Astronomy, 
                   Northwestern University,
                   Evanston, IL 60208,USA}

\author{JIANWEI QIU}

\address{Physics Department, 
                   Brookhaven National Laboratory, 
                   Upton, NY 11973\\
                   C.N. Yang Institute for Theoretical Physics, 
                   Stony Brook University, 
                   Stony Brook, NY 11794}

\maketitle


\begin{abstract}
We propose the use of 1-jettiness, a global event shape, for exclusive single jet production in lepton-nucleus deep inelastic scattering (DIS). We derive a factorization formula, using the Soft-Collinear Effective Theory,  differential in the transverse momentum and rapidity of the jet and the 1-jettiness event shape. It provides a quantitative measure of the shape of the final-state QCD radiation in the presence of the hard jet, providing a useful powerful probe of QCD and nuclear physics. For example, one expects differences in the observed pattern of QCD radiation between large and small nuclei and these can be quantified by the 1-jettiness event shape. Numerical results are given for this new DIS event shape at leading twist with resummation at the next-to-next-to-leading logarithmic (NNLL) level of accuracy, for a variety of nuclear targets. Such studies would be ideal at a future  EIC or LHeC electron-ion collider, where a range of nuclear targets are planned.

\end{abstract}

\ccode{PACS numbers: 11.25.Hf, 123.1K}

\section{Introduction}

It is well-known that global event shapes, such as thrust distributions at $e^+e^-$ colliders, have played a vital role in furthering our understanding of QCD.  The concept of event shapes for DIS was first introduced and developed\cite{Antonelli:1999kx,Dasgupta:2001sh,Dasgupta:2001eq,Dasgupta:2002bw} more than a decade ago. Thrust\cite{Antonelli:1999kx} and Broadening\cite{Dasgupta:2001eq} distributions were studied at the next-to-leading-log (NLL) level of accuracy and matched at ${\cal O}(\alpha_s)$ to  fixed order results. A numerical comparison was also done against ${\cal O}(\alpha_s^2)$ results\cite{Catani:1996vz,Graudenz:1997gv}. Thrust distributions have also been measured at HERA by the H1\cite{Adloff:1997gq,Aktas:2005tz,Adloff:1999gn} and ZEUS\cite{Breitweg:1997ug,Chekanov:2002xk,Chekanov:2006hv} collaborations.

\begin{figure}
\begin{center}
\includegraphics[scale=0.2]{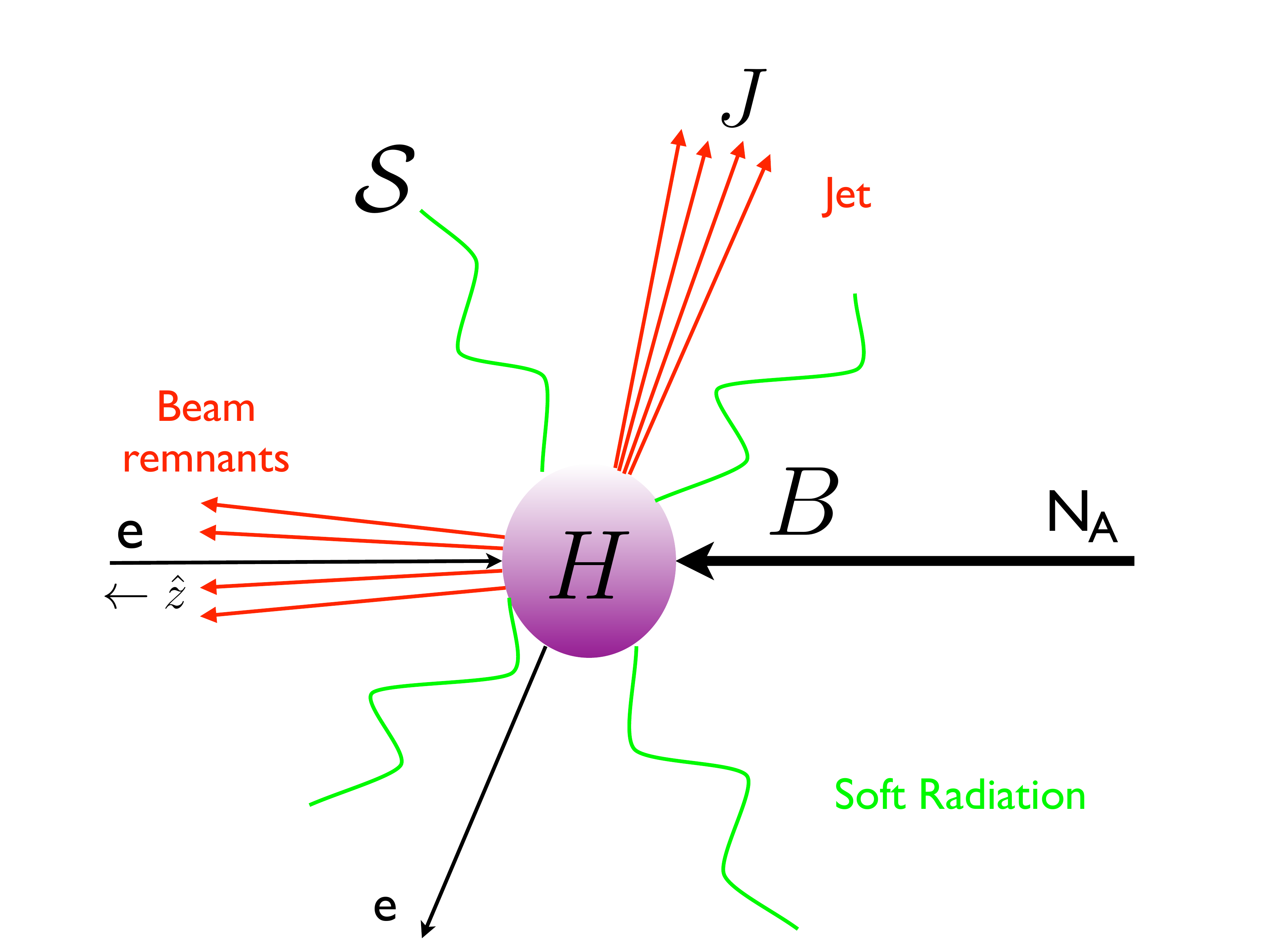}
\end{center}
\caption{Schematic figure of the process $e^- + N_A \to J + X$ in the limit $\tau_1\ll P_{J_T}$. The restriction $\tau_1\ll P_{J_T}$ allows only soft radiation of energy $E\sim \tau_1$ between the beam and jet directions. The factorization framework for this process is schematically shown in Eqs. (\ref{schem-1}) and (\ref{schem-2}).}
\label{fig:process}
\end{figure}

In this work we use N-Jettiness\cite{Stewart:2010tn}, a global event shape for exclusive N-jet production, to study single jet production in electron-nucleus collisions
\bea
\label{proc}
e^- + N_A \to  J + X,
\eea
where $N_A$ denotes a nucleus of atomic weight $A$ and $J$ denotes the leading final-state jet.  The additional hadronic radiation, contained in $X$, will be highly restricted (see Fig.\ref{fig:process}) for small values of the 1-jettiness ($\tau_1$) event shape. $\tau_1$ distributions provide theoretical control over the amount of radiation between the beam and jet directions, which can be used as a probe of QCD and nuclear medium effects. N-Jettiness\cite{Stewart:2010tn} was first introduced in the context of implementing jet vetoes at hadron colliders and has now been applied to several processes\cite{Stewart:2009yx,Stewart:2010pd,Liu:2012zg,Jouttenus:2013hs}. We adapt  this technology to exclusive jet production in deep-inelastic scattering (DIS).  

The use of 1-jettiness as a DIS event shape to probe of QCD and nuclear dynamics was first proposed in Ref.~\refcite{Kang:2012zr}.
In particular, the 1-jettiness event shape observable, for the process in Eq.(\ref{proc}), was defined as
\begin{equation}
\label{obs}
d\sigma_A \equiv \frac{d^3\sigma (e^- + N_A  \to J  + X)}{dy\> dP_{J_T}\>d\tau_1},
\end{equation}
where the cross-section is differential in $\tau_1$ and the transverse momentum ($P_{J_T}$) and rapidity ($y$) of the jet. This observable was explored in the region of phase space characterized by the restriction
\begin{equation}
\label{pscond}
\tau_1 \ll P_{J_T},
\end{equation}
which is effectively an inclusive veto on additional jets.
A typical event corresponding to this region of phase space is schematically shown in Fig. \ref{fig:process}. It corresponds to a single narrow jet with only soft radiation of energy $E\sim \tau_1 \ll P_{J_T}$ allowed between the beam and jet directions. The physics of this region of phase space is dominated by collinear emissions along the beam and jet  directions and soft emissions throughout the event. Large Sudakov logarithms appear of the form $\alpha_s^{n}\ln^{m}(P_{J_T}/\tau_1)$ for $m\leq 2n$ and correspond to the jet-veto logarithms. The Soft-Collinear effective theory (SCET) \cite{Bauer:2000ew,Bauer:2000yr,Bauer:2001ct,Bauer:2001yt,Bauer:2002nz,Beneke:2002ph},  is the appropriate effective theory for this region of phase space and facilitates a resummation of the Sudakov logarithms within a factorization framework.

Such a factorization framework for the process in Eq.(\ref{obs}), in the region $\tau_1 \ll P_{J_T}$, was first derived in Ref.~\refcite{Kang:2012zr}. Numerical results were presented for a proton target at the NLL level of accuracy. In Ref.~\refcite{Kang:2013wca}, results were extended to include resummation at next-to-next-to-leading logarithmic (NNLL) level of accuracy, for a variety of nuclear targets. This was the first time that NNLL resummation was achieved for a DIS event shape. Subsequently, 1-jettiness for DIS was studied in Ref.~\refcite{Kang:2013nha} and they also independently presented results at NNLL level of accuracy for the proton target. They studied three different versions of 1-jettiness for DIS which they denoted as $\tau_a,\tau_b,$ and $\tau_c$. These different versions correspond to different choices for the reference vectors, used to define the 1-jettiness event shape, and have correspondingly different factorization structures as presented in Ref.~\refcite{Kang:2013nha}. The event-shape $\tau_a$ is equivalent to  $\tau_1$ first studied in Refs.~\refcite{Kang:2012zr,Kang:2013wca}, $\tau_b$ was shown to be equivalent to the thrust distribution studied in Ref. \refcite{Antonelli:1999kx}, and  $\tau_c$  is a new definition of 1-jettiness that is naturally conducive for analysis in the target rest frame. Together with the works of Refs.~\refcite{Antonelli:1999kx,Dasgupta:2001sh,Dasgupta:2001eq,Dasgupta:2002bw,Kang:2012zr,Kang:2013wca,Kang:2013nha}, a large class of  DIS event shapes have now been explored. These works can be viewed as complementary to each other, as they provide independent quantitative  measures of the properties of the final-state QCD radiation in DIS processes.

\section{Formalism}
Here we focus on the work of Refs.~\refcite{Kang:2012zr,Kang:2013wca}. The cross-section in Eq.(\ref{obs}) is computed in the center of mass frame (see Fig.~\ref{fig:process}) defined by the electron momentum and the \textit{average} nucleon momentum in the nucleus. The momentum of the incoming electron ($p_e$) and the nucleus ($P_A$) can then be written as
$p_e^\mu = (p_e^0,\vec{p}_e), \>P_A^\mu = A \> (p_e^0,-\vec{p}_e)$
so that the electron and nucleus are treated as massless $p_e^2=0$. By setting $p_e^0=|\vec{p_e}|=Q_e/2$, the center of mass energy is given by
\bea
s=(p_e +P_A)^2 = A Q_e^2.
\eea
The 1-jettiness event shape is defined as
\begin{equation}
\label{1-jettiness}
\tau_1 = \sum_k \text{min} \Big \{ \frac{2q_A\cdot p_k}{Q_a}, \frac{2q_J\cdot p_k }{Q_J}\Big \},
\end{equation}
where the sum is over all final state particles (except the lepton) with momenta denoted by $p_k$. The null vectors $q_A$ and $q_J$ denote reference vectors along the nuclear beam and jet directions respectively. The parameters $Q_a$ and $Q_J$ are on the order of the hard scale $ P_{J_T}$ and different choices correspond to different definitions of $\tau_1$. The reference vectors $q_A$ and $q_J$ can be determined via a minimization\cite{Thaler:2011gf} condition such that the optimal choice minimizes the value of $\tau_1$. In this procedure, the analysis can be performed without reference to any jet algorithm and is analogous to finding the thrust axis in the calculation of thrust distributions. One can also choose $q_A$ to align with the beam axis and determine $q_J$ via a standard jet algorithm. Note that in this case, the only information from the jet algorithm that is used in the calculation of $\tau_1$ is the corresponding determination of the reference vector $q_J$. $\tau_1$ has no explicit dependence on other jet algorithm parameters such as the jet  radius R. This property of the event shape formalism allows for better analytic control over higher order perturbative corrections.

From Eq.(\ref{1-jettiness}), we see that energetic radiation at wide angles from the beam ($q_A$) and jet ($q_J$) directions make the largest contributions to $\tau_1$. Thus, by restricting to the region  $\tau_1 \ll P_{J_T}$, energetic radiation ($E\sim P_{J_T}$) is only allowed along the beam and jet directions. This corresponds to the picture shown schematically in Fig.\ref{fig:process}, where the radiation at wide angles from the beam and jet directions is restricted to be soft ($E\sim \tau_1 \ll P_{J_T}$). Thus, in this region the dependence of any jet algorithm will be associated with how soft radiation at wide angles is grouped into the jet. This effect in determining the reference vector $q_J$ is thus power suppressed in $\tau_1/P_{{J_T}}$. 

For our numerical analysis, we choose the reference vectors $q_A$ and $q_J$ as
\bea
\label{refvecJ}
q_A^\mu = x_A P_A^\mu, \qquad q_J^\mu=P_J^\mu = (P_{J_T}\cosh y, \vec{P}_{J_T},P_{J_T}\sinh y),
\eea
where  $x_A$ is the nucleus momentum fraction carried by the incoming parton that enters the hard interaction, and $P_J$ is the total momentum of the final-state jet. Thus, the reference vector $q_J$ is simply defined as the massless vector constructed from the differential quantities $P_{J_T}$ and $y$ in the cross-section in Eq.(\ref{obs}). For the parameters $Q_a$ and $Q_J$ we choose
\begin{equation}
\label{choices}
 Q_a=x_A A Q_e, \qquad  Q_J = 2P_{J_T}\cosh y.
\end{equation}
The total jet momentum is defined as 
\begin{equation}
\label{pjet}
P_J = \sum_k p_k \>\theta (\frac{2q_A\cdot p_k}{Q_a} - \frac{2q_J\cdot p_k}{Q_J} ).
\end{equation}
This definition is closely related to 1-jettiness. From Eq.(\ref{1-jettiness}), it is clear that each  final state hadronic particle of momentum $p_k$ is associated either with the beam ($q_A$) or jet $(q_J)$ directions through minimization condition in Eq.(\ref{obs}). The total jet momentum is defined as the sum of the momenta of all particles associated with the $q_J$ direction, as quantified by the step function in Eq.(\ref{pjet}). In the region $\tau_1 \ll P_{J_T}$, the total jet momentum (and correspondingly $q_J$) obtained from a different jet algorithm, will differ by power corrections in $\tau_1/P_{J_T}$ associated with how wide-angle soft radiation is grouped into the jet.
\begin{figure}
\subfigure [$R_u^{\text{Ur(V)}}$] { \label{fig:subfig1}\includegraphics[scale=0.3]{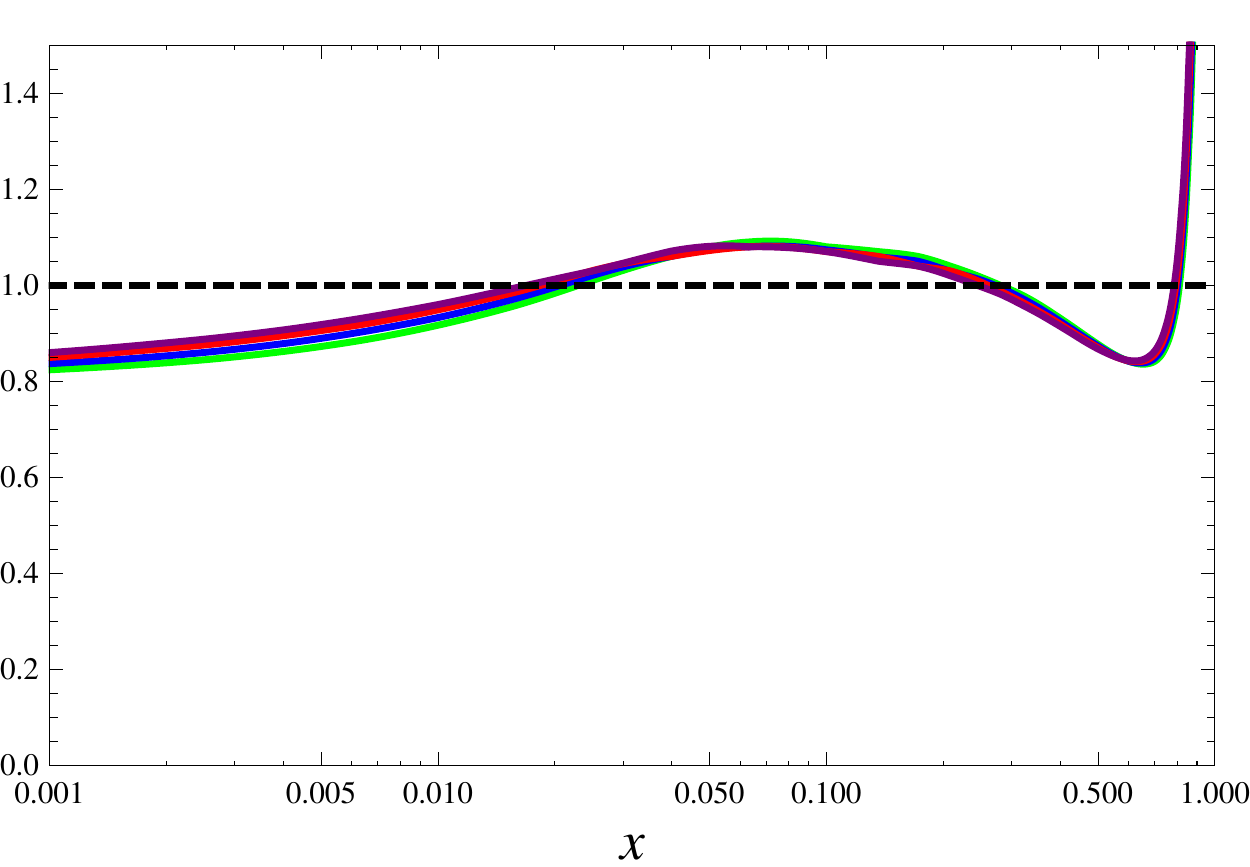}}
\subfigure [$R_d^{\text{Ur(V)}}$] { \label{fig:subfig2}\includegraphics[scale=0.3]{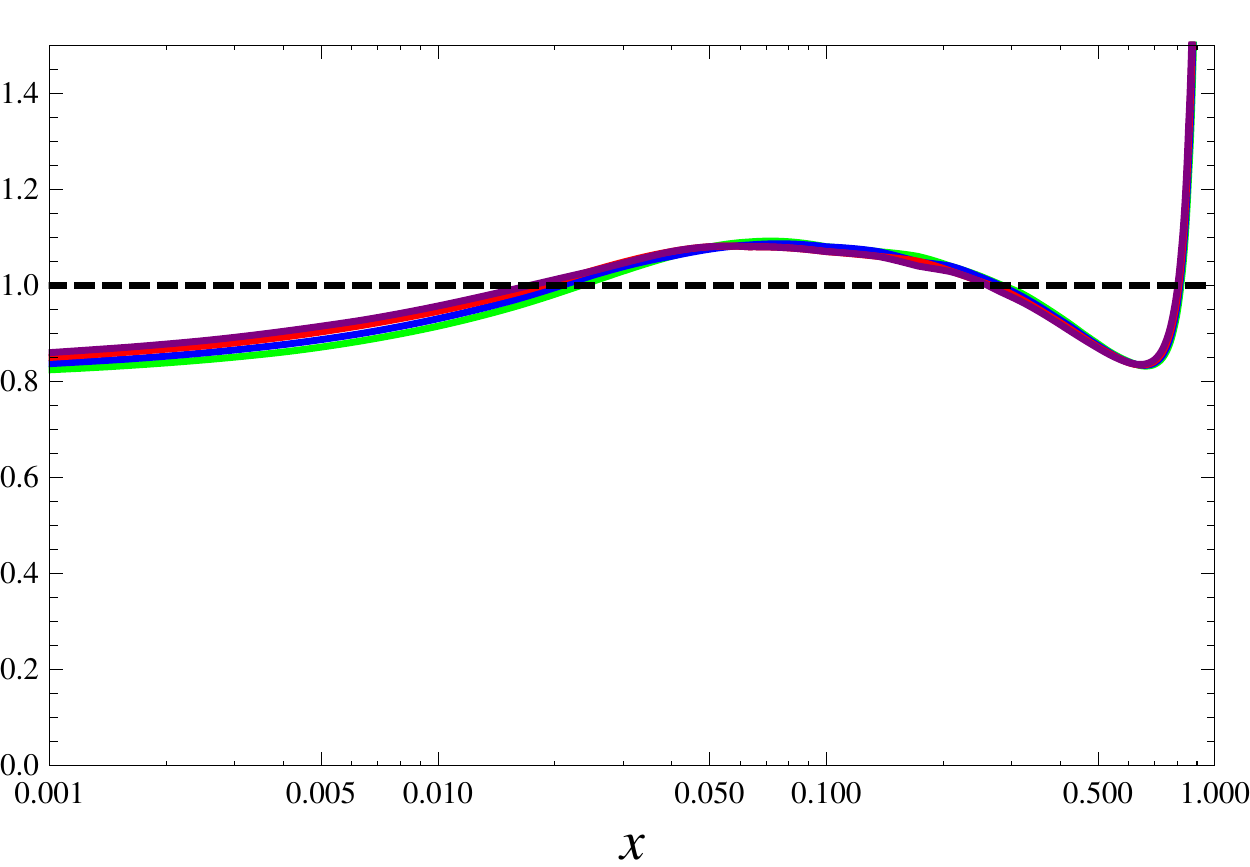}}
\subfigure [$R_u^{\text{Ur(S)}}$] { \label{fig:subfig3}\includegraphics[scale=0.3]{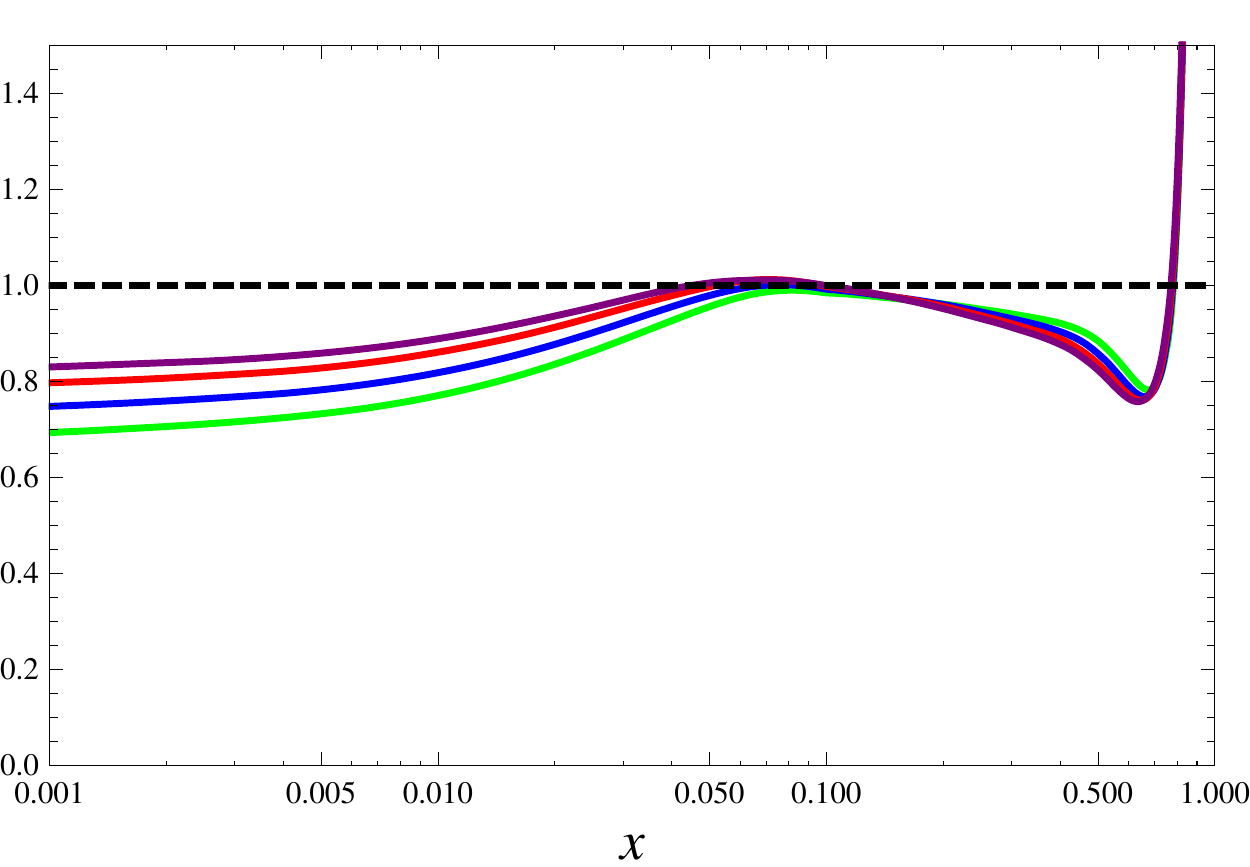}}
\subfigure [$R_d^{\text{Ur(S)}}$] { \label{fig:subfig4}\includegraphics[scale=0.3]{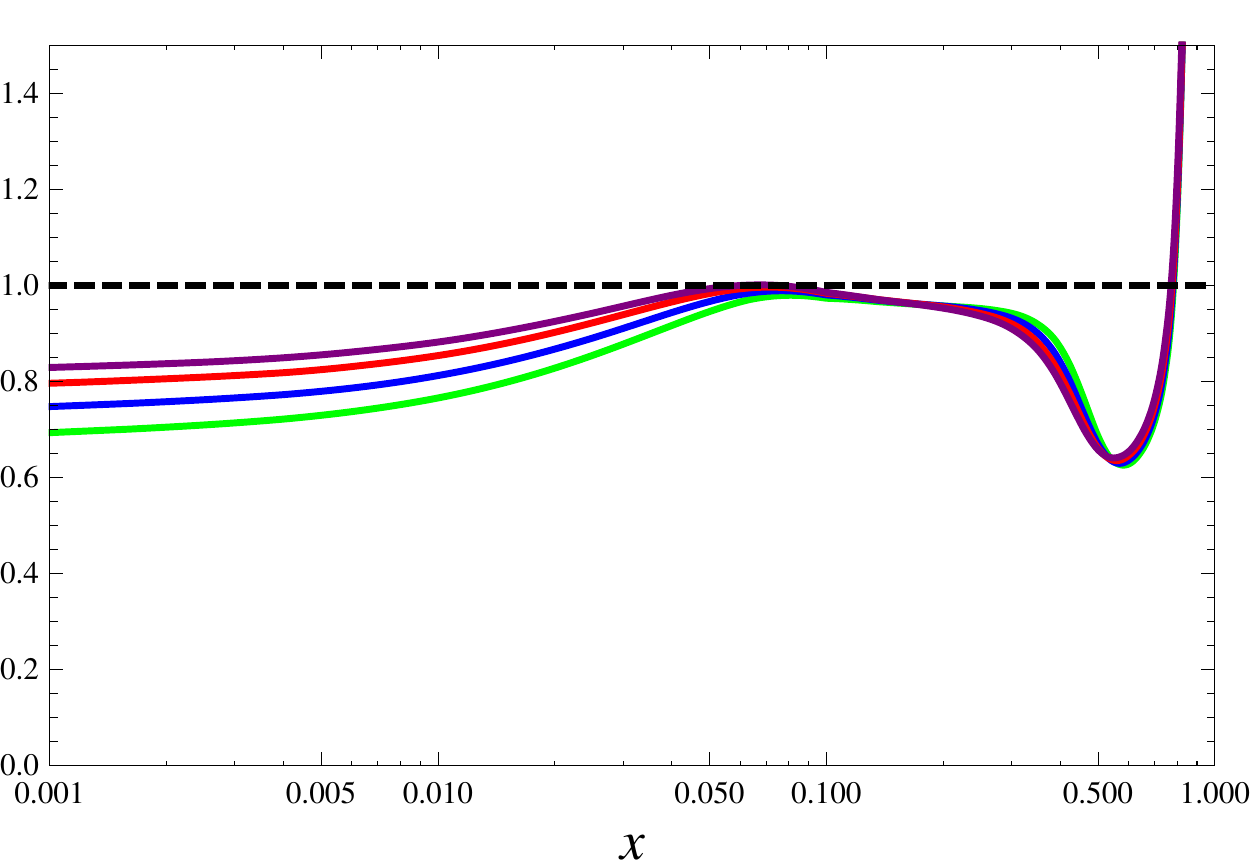}}
\subfigure [$R_s^{\text{Ur}}$] { \label{fig:subfig5}\includegraphics[scale=0.3]{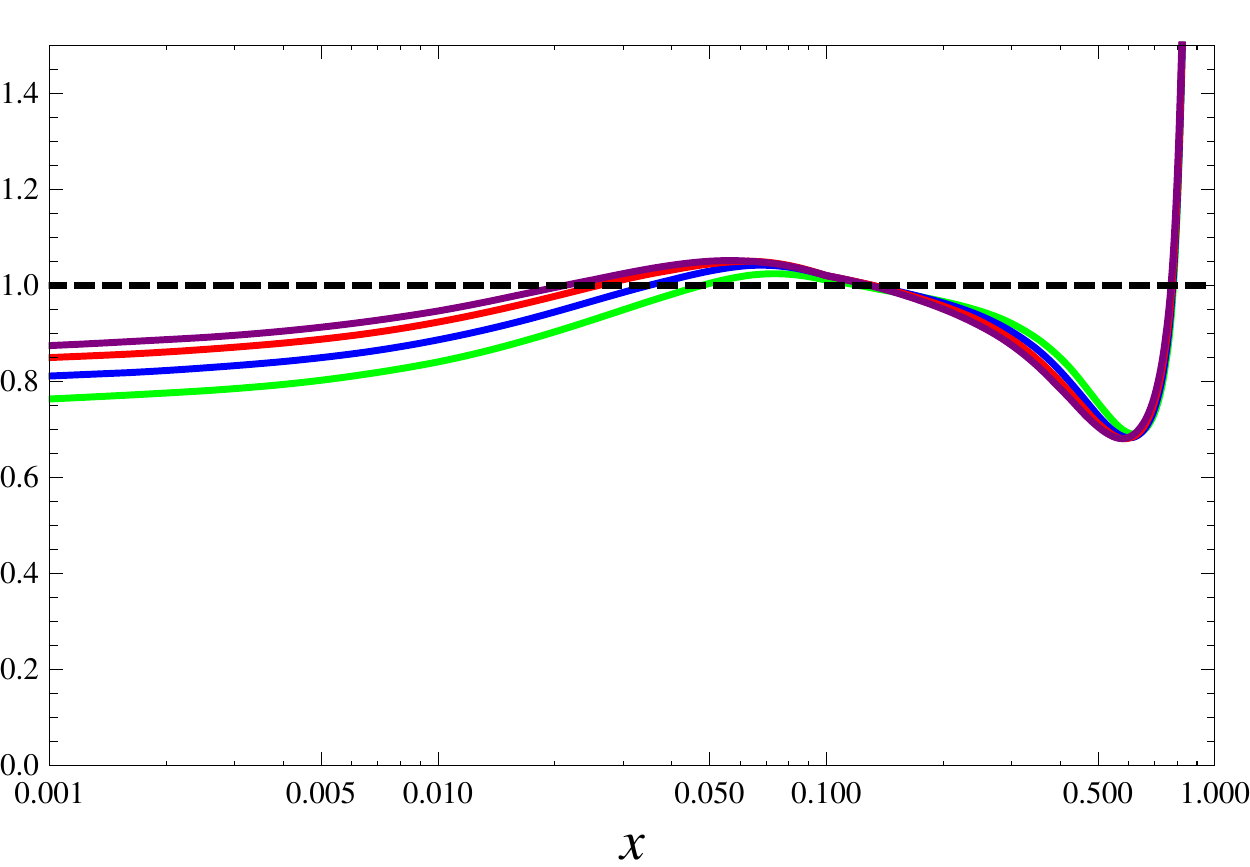}}
\subfigure [$R_g^{\text{Ur}}$] { \label{fig:subfig8}\includegraphics[scale=0.3]{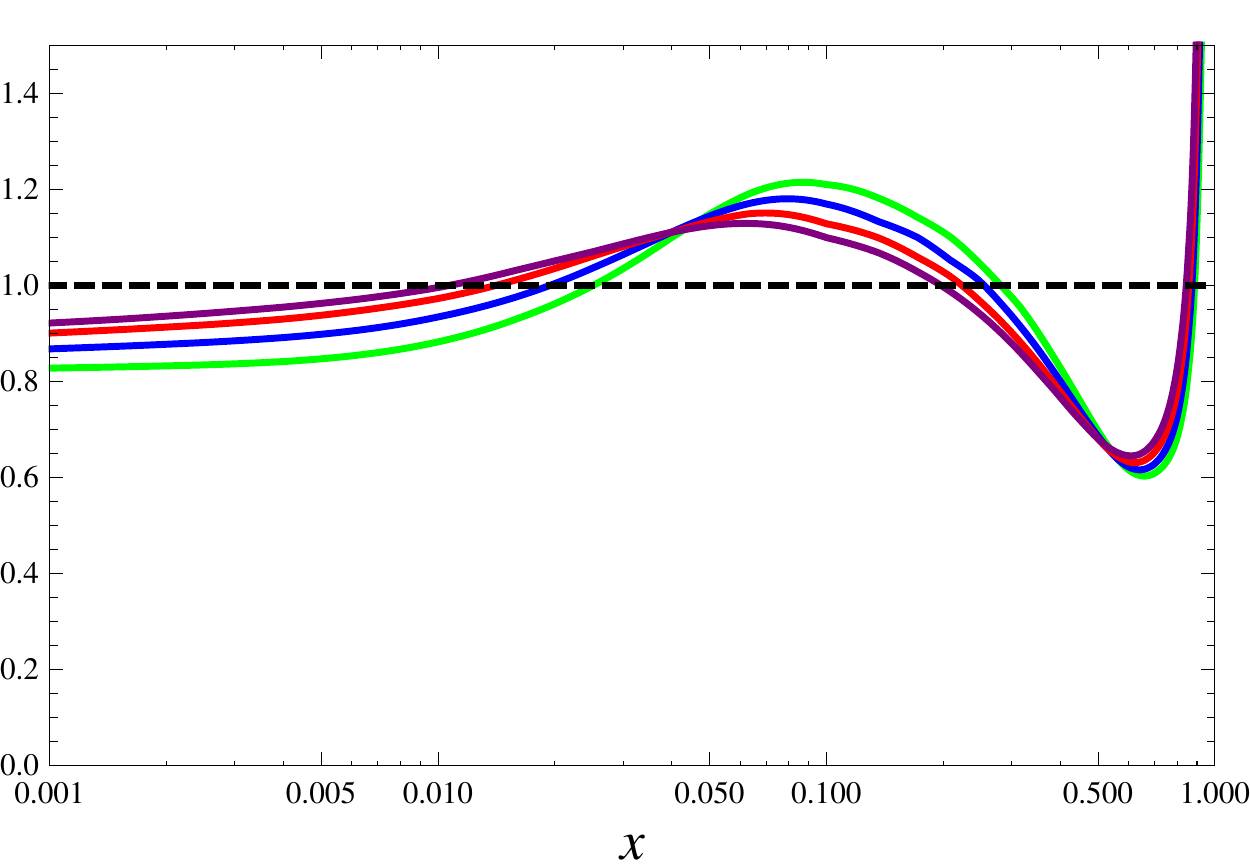}}
\caption{Nuclear correction factors $R_i^{\text{Ur}}(x,\mu)$ for the NLO nuclear PDF for a Uranium target as defined in Eq.(\ref{EPS09}). The subscript $i$ runs over the parton species $i=\{u,d,s,g\}$. For the $u$ and $d$ quarks, separate  $R$-factors are given for the valence (V) and sea quarks (S). The different curves in each graph correspond to different values for the scale $\mu$. By looking at the region of small Bjorken-$x$, the different curves from the bottom to the top correspond to $\mu=3$ GeV (Green), $\mu=5$ GeV (Blue), $\mu=10$ GeV (Red), and $\mu=20$ GeV (Purple). These plots were generated using publicly available code for the EPS09 PDF set.}
\label{Rfactors}
\end{figure}

The factorization formula\cite{Kang:2012zr,Kang:2013wca} for the observable in Eq.(\ref{obs}) has the schematic form
\bea
\label{schem-1}
\frac{d^3\sigma}{dy dP_{J_T} d\tau_1} \sim H \otimes B \otimes J \otimes {\cal S},  
\eea
where $H,B,J,$ and $S$ (see Fig.~\ref{fig:process}) denote the hard, nuclear beam, jet, and soft functions respectively. The hard function is associated with the hard partonic interaction that produces the final-state jet. The beam function\cite{Stewart:2009yx} captures the physics of the initial state parton distributions and correlations in the nucleus, energetic ($E\sim P_{J_T}$) initial-state collinear radiation, and beam remnants. The jet function describes the dynamics of collinear radiation along the jet direction. Finally, the soft function describes the dynamics of soft ($E\sim \tau_1$) radiation throughout the event. All of these objects have well-defined field-theoretic definitions and can be found listed in Ref.~\refcite{Kang:2013wca}. The beam function is sensitive to two disparate scales associated with the perturbative emissions of initial-state radiation and the non-perturbative physics of the initial state nucleus. The physics of these two scales can be separated\cite{Stewart:2009yx} through an operator product expansion (OPE) so that at leading twist, the beam function can be written as a convolution between a perturbatively calculable coefficient ${\cal I}$ and the standard nuclear parton distribution function (PDF) ($f_A$) 
\begin{equation}
\label{schem-2}
B \sim {\cal I} \otimes f_A  + {\cal O}\Big [ \frac{Q_s^2 (A)}{\tau_1 P_{J_T}} \Big ] .
\end{equation}
The coefficient ${\cal I}$ describes the physics of the perturbative initial-state collinear emissions that knocks the initial parton off-shell by an amount $p^2 \sim \tau_1 P_{J_T} $. Power corrections in the OPE depend on the nuclear scale $Q_s$ which depends on the nuclear species. The dependence of $Q_s$ on the atomic weight of the nucleus is typically parameterized as
\begin{equation}
\label{nucscale}
Q_s^2(A) \sim A^\alpha \Lambda_{QCD}^2,
\end{equation}
where the parameter $\alpha$ depends on the details of the underlying nuclear physics. Thus, one can explore nuclear-dependent power corrections in the space of $\{ A, P_{J_T}, \tau_1\}$, by looking for deviations from the predictions of the leading twist formula.

\section{Numerical Results}
\begin{figure}
\begin{center}
\includegraphics[scale=0.5]{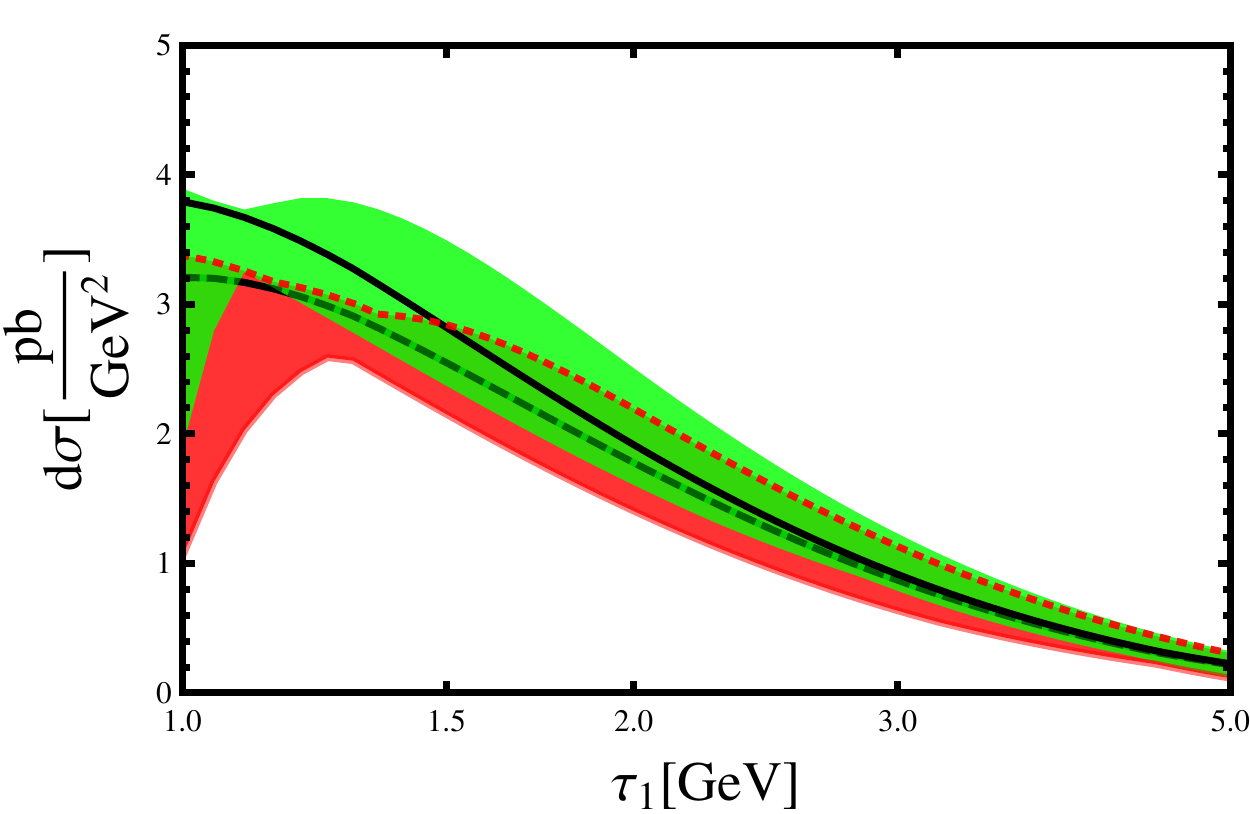}
\end{center}
\caption{$\tau_1$-distribution for a proton target with NLL$^\prime$ (lower red band) and NNLL (upper green band) resummation for $Q_e=90$ GeV, $P_{J_T}=20$ GeV and $y=0$.}
\label{protontau1}
\end{figure}

The more detailed version of the schematic factorization formulae in Eqs.(\ref{schem-1}) and (\ref{schem-2}) is given by
\begin{eqnarray}
\label{factorization-EPS09}
d\sigma_A(\tau_1,P_{J_T},y)\equiv \frac{d^3\sigma}{dy dP_{J_T} d\tau_1}\Big |_{\text{EPS09}} &=&\sigma_0 \sum_{q,i} e_q^2 \int_{x_*}^{1} \frac{dx}{x} \int ds_J \int dt_a  
 \nonumber\\
&&\times H(\xi^2, \mu; \mu_H)J^q(s_J, \mu;\mu_J){\cal I}^{qi}\left(\frac{x_*}{x}, t_a, \mu;\mu_B\right)
\nonumber \\
&&\times {\cal S}\left(\tau_1 - \frac{t_a}{Q_a}-\frac{s_J}{Q_J}, \mu;\mu_S\right) f_{i/A}^{EPS09}(x,\mu_B), \nonumber  \\
\end{eqnarray}
where we have used the EPS09\cite{Eskola:2009uj} nuclear PDF set for generating numerical results. This is the master formula used for generating all numerical results at leading twist with NNLL resummation. For more details about this formula we refer the reader to the original paper in Ref.~\refcite{Kang:2013wca}. Note that all of the nuclear dependence in the factorization formula is contained entirely in the nuclear PDFs $f_{i/A}$. All other objects are independent of the nuclear target giving rise to universality among processes with different nuclear targets. The nuclear PDFs are parameterized  as
\begin{figure}
\subfigure { \label{fig:subfig1}\includegraphics[scale=0.45 ]{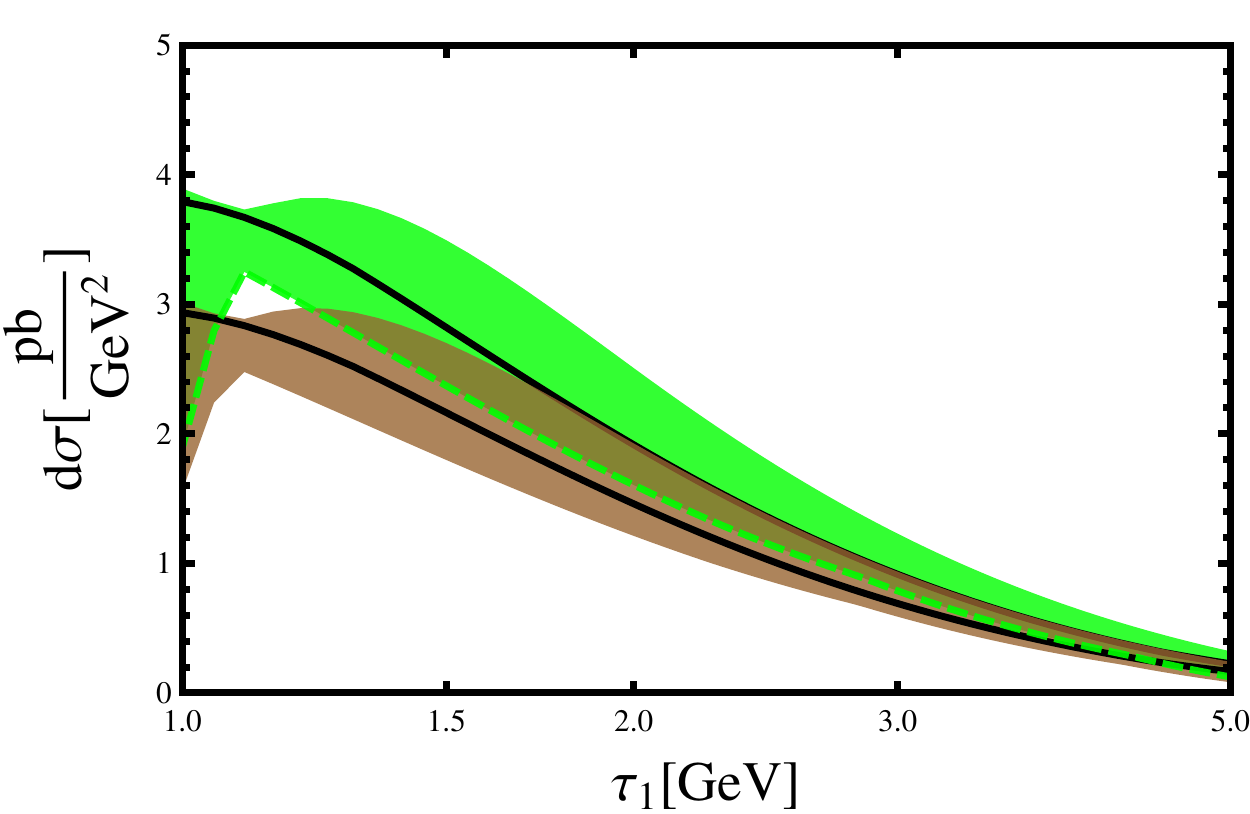}}
\subfigure { \label{fig:subfig2}\includegraphics[scale=0.45]{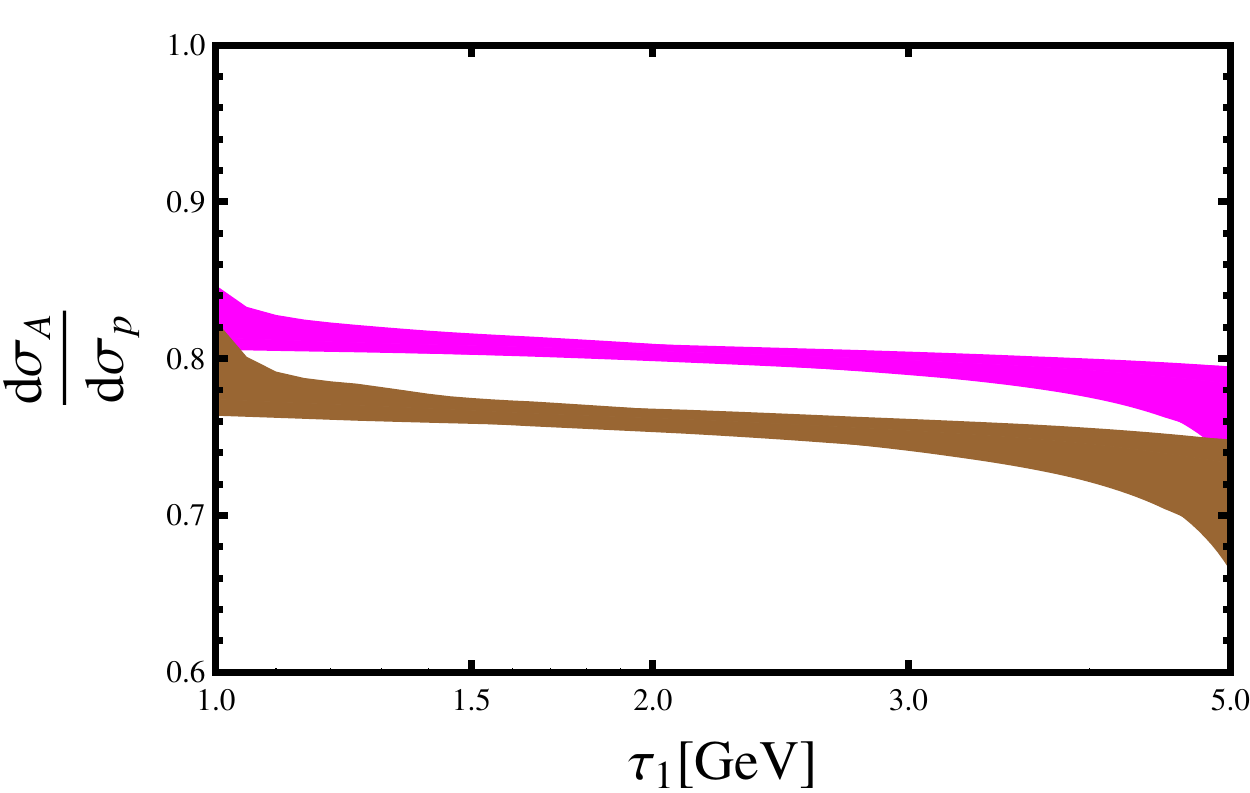}}
\caption{Left Panel: $\tau_1$-distributions with NNLL resummation for the proton (upper green brand) and Uranium (lower brown band) targets at $Q_e=90$ GeV, 
 $P_{J_T}=20$ GeV, and $\tau_1=1.5$ GeV. Right Panel: Ratio of Uranium to proton (lower brown band) and Carbon to proton (upper magenta band) $\tau_1$-distributions at the same kinematics.}
\label{tau1nuc}
\end{figure}
\begin{eqnarray}
\label{EPS09}
f_{u/A}^{EPS09}(x,\mu) &=&   \frac{Z}{A} \>R_u^A (x,\mu) \>f_{u/p}(x,\mu) + \frac{A-Z}{A} \>R_d^A(x,\mu) \>f_{d/p}(x,\mu) , \nn \\
f_{d/A}^{EPS09}(x,\mu) &=& \frac{Z}{A}\> R_i^A (x,\mu)\> f_{d/p}(x,\mu) + \frac{A-Z}{A}\> R_u^A(x,\mu) \> f_{u/p}(x,\mu) ,\nn \\
f_{{s,c,b}/A}^{EPS09}(x,\mu) &=& R_{s,c,b}^A(x,\mu)\> f_{{s,c,b}/p}(x,\mu), \nn \\
f_{g/A}^{EPS09}(x,\mu) &=& R_g^A(x,\mu)\> f_{g/p}(x,\mu),
\end{eqnarray}
where $Z$ denotes the atomic number of the nucleus, the $f_{i/p}$ denote the standard proton PDFs, and the $R_{i}^A$ denote nuclear correction factors. Isospin symmetry has been used to write neutron PDFs in terms of proton PDFs. The nuclear correction factors parameterize  well-known nuclear effects. Shadowing effects suppress the nuclear PDFs at small-$x$.  On the other hand, anti-shadowing gives rise to an enhancement of the nuclear PDFs in the region $x\sim 0.1$. The EMC effect suppresses the parton density for moderate values $x> 0.2$ and the Fermi motion of nucleons causes an enhancement for large values of $x$. These effects are illustrated in Fig.~\ref{Rfactors} where nuclear correction factors for various parton densities are shown for the Uranium nucleus. 
As seen in Eq.(\ref{factorization-EPS09}), the nuclear PDFs are integrated over the range  $[x_*,1]$  where 
\bea
\label{xstar}
x_* &=& \frac{e^y P_{J_T}}{Q_e-e^{-y}P_{J_T}}.
\eea
Thus, by exploring the kinematic space spanned by $\{Q_e, P_{J_T}, y \}$, one can gain sensitivity to different regions of Bjorken-$x$ and probe the various nuclear effects.

In Fig.~\ref{protontau1}, we show the $\tau_1$ distribution for a proton target at $Q_e=90$ GeV, $P_{J_T}=20$ GeV, and $y=0$. We show numerical results for NLL+NLO, denoted as NLL$'$ (lower red band), and NNLL resummation (upper green band).  The resummation of the Sudakov logs of $\tau_1/P_{J_T}$ tames the singular behavior of the fixed order cross-section as one makes the jet-veto more and more restrictive by going to the region of small $\tau_1$. The width of the bands estimate the perturbative uncertainty from higher order effects and are obtained by standard scale variation procedures. For $\tau_1\sim \Lambda_{QCD}$, the universal soft function becomes non-perturbative must be modeled. For more details and numerical results for the region $\tau_1\sim \Lambda_{QCD}$, we refer the reader to Ref.~\refcite{Kang:2013wca}. 

In the left panel of Fig.~\ref{tau1nuc} we show the $\tau_1$ distribution with NNLL resummation for the proton (upper green band) and Uranium (lower brown band). In the right panel we show the ratio of distributions. The lower brown band shows the ratio of the Uranium to proton $\tau_1$ distribution. We also show the ratio of Carbon to proton distribution as the upper purple band. Note that there is a dramatic reduction in the scale variation undertainties in the ratio of $\tau_1$ distribution between different nuclear targets. These ratios also deviate from unity well outside the theoretical uncertainty bands showing that these distributions are effective probes of nuclear effects. 
\begin{figure}
\subfigure [$\>$ C and Ur ] { \label{fig:subfig1}\includegraphics[scale=0.45]{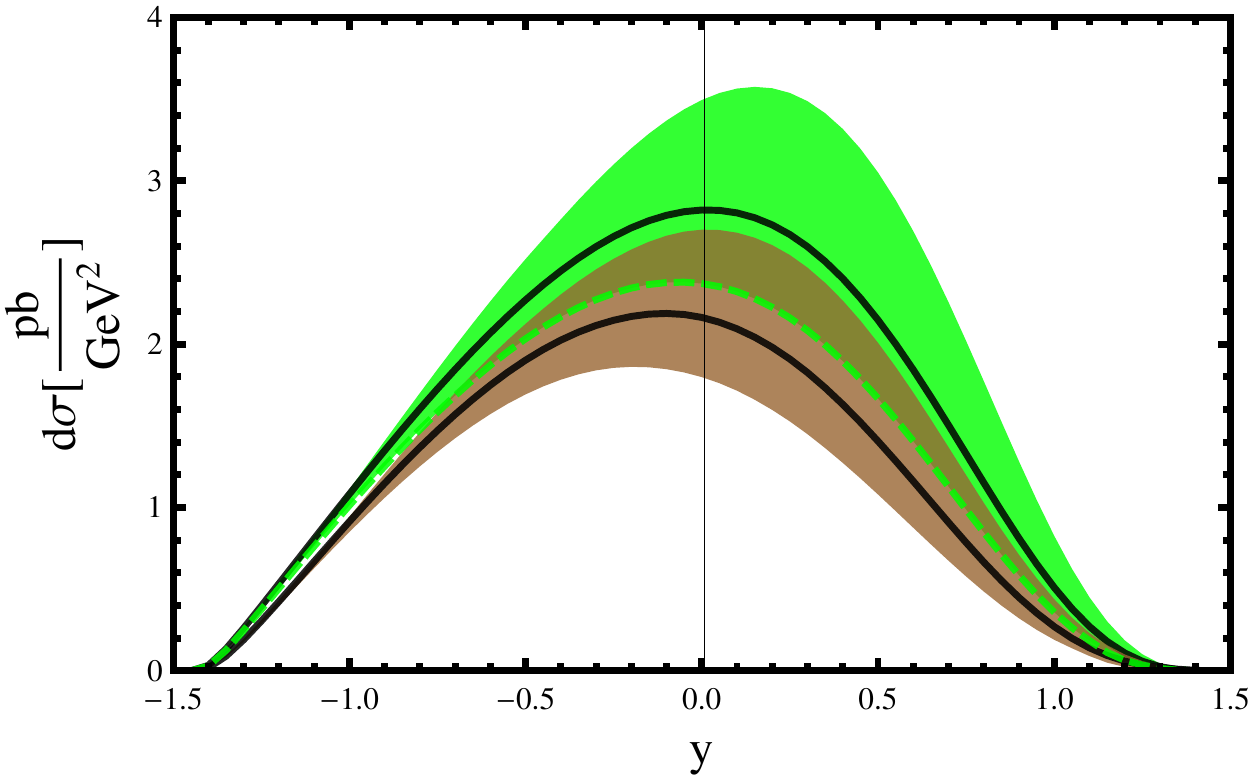}}
\subfigure [$\>$ C and Ur ]{ \label{fig:subfig2}   \includegraphics[scale=0.45]{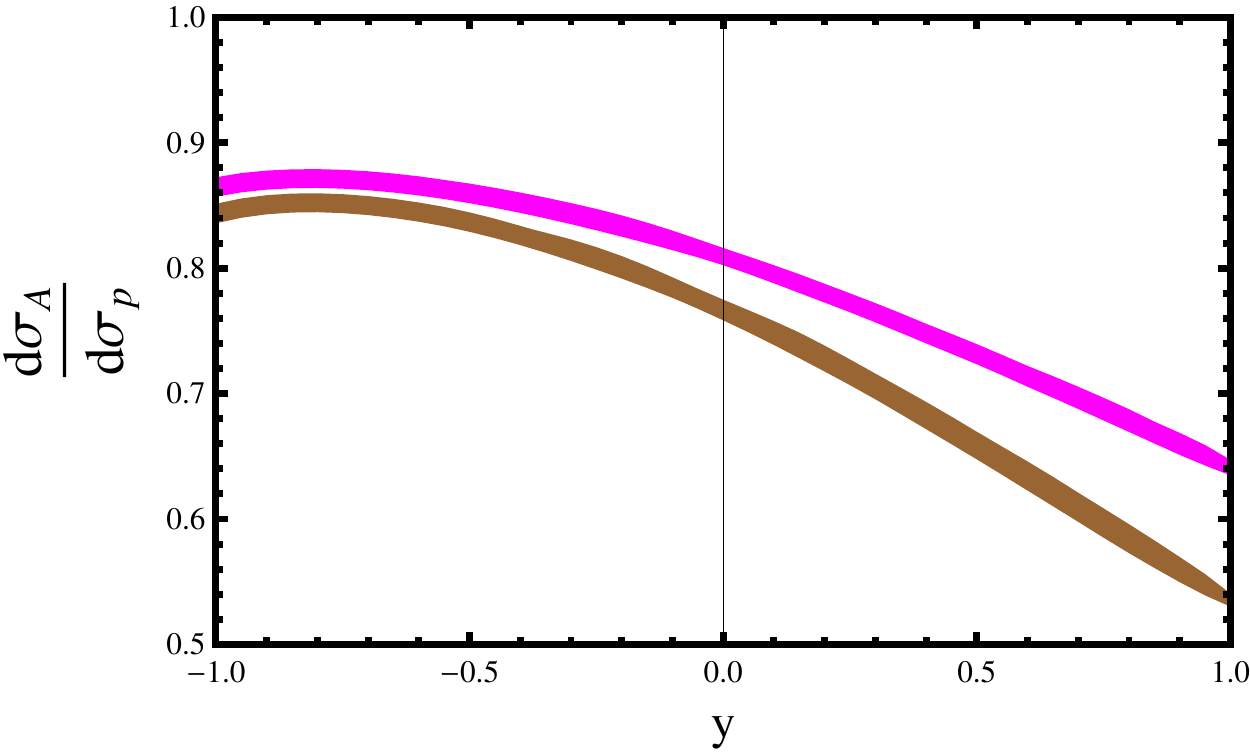}}
\caption{Left Panel: Rapidity ($y$) distributions with NNLL resummation for the proton (upper green brand) and Uranium (lower brown band) targets at $Q_e=90$ GeV, 
 $P_{J_T}=20$ GeV, and $\tau_1=1.5$ GeV. Right Panel: Ratio of Uranium to proton (lower brown band) and Carbon to proton (upper magenta band) rapidity distributions at the same kinematics.}
 \label{rapidity}
\end{figure}

In the left panel of Fig.~\ref{rapidity} we show the rapidity distributions with NNLL resummation for the proton (upper green band) and Uranium (lower brown band) targets at the fixed values of $\tau_1=1.5$ GeV, $Q_e=90$ GeV, and $P_{J_T}=20$ GeV. In the right panel we show the ratio of the Uranium to proton rapidity distributions (lower brown band). We also show the ratio of the Carbon to proton distributions. Once again, there is a dramatic reduction in the scale variation uncertainty in the ratio of distributions. In addition the ratio of rapidity distributions of different nuclear targets has an interesting shape corresponding to the fact that different values of rapidity probe different regions of Bjorken-$x$ in the nuclear PDFs, as seen from Eq.(\ref{xstar}). 

We refer the reader to Ref.~\refcite{Kang:2013wca} for more detailed discussions and many more numerical results that include a large variety of nuclear targets, distributions in the other kinematic variables $P_{J_T}$ and $Q_e$, and distributions in the non-perturbative soft region $\tau_1\sim \Lambda_{QCD}$. 

\section*{Acknowledgements}

This work was supported in part by  the U.S. Department of Energy under contract numbers  DE-AC02-05CH11231 (ZK), DE-AC02-98CH10886 (JQ), DE-AC02-06CH11357 (XL) and the grants  DE-FG02-95ER40896 (XL) and DE-FG02-08ER4153 (XL), and the U.S. National Science Foundation under grant NSF-PHY-0705682 (SM).

\end{document}